\documentstyle[11pt,aaspp4]{article}

\lefthead{Delgado et al.}
\righthead{NGC 2362}

\begin{document}

\title{Multiwavelength analysis of the Young Open Cluster NGC 2362}

\author{A. J. Delgado, O. Gonz\'alez-Mart\'\i n, E.J. Alfaro}
\affil{Instituto de Astrof\'\i sica de Andaluc\'\i a,
CSIC, Apdo 3004, 18080-Granada, Spain}

\and

\author{J. Lin Yun}
\affil{Observat\'orio Astron\'omico de Lisboa,
Tapada da Ajuda
1349-018 Lisboa, Portugal}




\begin{abstract}
We present a multiwavelength analysis of the young open cluster NGC 2362. 
$UBVR_{C}I_{C}$ CCD photometric observations, together with available data 
in the {\sl Chandra} data base, near infrared data from the Two Micron 
All Sky Survey (2MASS), and recently published H$\alpha$ spectroscopy 
were used to get information about the evolutionary stage of the cluster 
and the main physical properties of its stellar content. Cluster 
membership is estimated for every individual star by means of ZAMS and 
isochrone fitting. The cluster is confirmed to host a richly populated 
pre-main sequence (PMS), and to contain a large amount of X-ray emitting 
stars, which reach from the PMS members of GK spectral type, up to the most 
luminous OB type main sequence (MS) members. The PMS cluster members 
show no significant age spread, and the comparison to both PMS and post-MS 
isochrones suggests a younger age for the more massive MS than for lower mass 
PMS members. The analysis allows to asses the validity of currently used 
pre-main sequence evolutionary models, and supports the suggestion of a well 
defined positive correlation of the X-ray emission from PMS stars with their 
bolometric luminosity. Clear differences are found on the other hand, between 
the X-ray activity properties of MS and PMS cluster members, both in the 
relation between X-ray luminosity and bolometric luminosity, and in spectral 
properties as well.
\end{abstract}




%

\section{Introduction}

The young open cluster NGC 2362 has been the subject of recent attention from 
several authors, as an adequate object for the study of the star formation 
processes (Moitinho et al 2001, Haisch et al. 2001, Dahm 2005, D05 in the 
following). Located in the third galactic quadrant, it is little affected by 
reddening and, despite its youth, shows a relative absence of intracluster 
material (Balona \& Laney 1996). All this
allows the observation of its stellar population in a wide range of
masses, moving from the low mass PMS stars to massive OB stars
populating the upper part of the color-magnitude diagram (CMD)

The photometric observations of NGC 2362 were carried out in the framework of
a current project devoted to detect and study  pre-main sequence (PMS) stars
among the members of young open clusters. This project is based on optical
$UBVRI$, and eventually H$\alpha$ observations of galactic clusters, inside 
the age range between 1 and 10 Myr, and located at distances from the Sun not 
farther than 3-4 kpc. These constraints should leave objects with observable 
PMS members of spectral types from A to K, detectable in photometric diagrams 
deep dwon to $V$ 21-22, depending on reddening. These observations are 
feasible with small telescopes, and can be obtained for a 
wide sample of clusters in the Cygnus and Perseus galactic spiral arms, as 
well as for some clusters located at larger distances in the direction of the 
galactic anticenter. A presentation of the results obtained up to now in this 
project will be the subject of another paper.

On the other hand, the investigation of X-ray emission from PMS stars received
increasing attention in recent years after the new space missions, able to 
provide measurements of high spectral and spatial resolution. As a 
consequence, the debate about the physical mechanisms that originate this 
activity has gained in richness and insight (see Preibisch et al. 2005, and 
references there in). 

In this paper we collect $UBVR_{C}I_{C}$ CCD photometry of our own, X-ray data
on sources detected in the field by the {\sl Chandra} Advanced CCD Imaging 
Spectrometer (ACIS)\footnote{{\tt http://cxc.harvard.edu/cda/}}, $JHK$ 
photometry from the 2MASS\footnote{{\tt http://www.ipac.caltech.edu/2mass/}} 
data base, and H$\alpha$ emission and Li absorption from PMS cluster members 
(D05), in order to analyze the evolutionary stage of the cluster members ant 
its connection with the X-ray activity.

\section{The data}

The optical observations were secured during two nights in December 2000, at 
the CTIO observatory with the YALO 1m telescope\footnote{YALO is the 
{\bf Y}ale-{\bf A}URA-{\bf L}isbon-{\bf O}hio consortium (Bailyn et al. 1999)}.
As mentioned above, these observations are included in a long term search for 
PMS stars in young open clusters. A thorough presentation of the southern 
clusters observed in this program, including the detailed description of the 
reduction and calibration procedures, will be the subject of a forthcoming 
paper. This publication will also include the photometric catalogue of all 
objects observed.

The pixel coordinates of our catalogue were transformed to equatorial 
coordinates using the IRAF\footnote{The Image Reduction and Analysis Facility 
(IRAF) is distributed by the national Optical Astronomical Observatory, which 
is operated by the Association of Universities for Research in Astronomy, Inc.
(AURA) under cooperative agreement with the National Science Foundation} tasks
{\tt ccmap} and {\tt xy2rd}. The matching provides $UBVR_{C}I_{C}JHK$ 
photometry for 725 stars. Among them, 551 have Photometric Quality flag A to D
in all three $JHK$ bands (see the catalogue description at 
{\tt http://www.ipac.caltech.edu/2mass/releases/allsky/doc/}).

Data from {\sl Chandra}-ACIS in the field of NGC~2362 have been retrieved from
the public data archive. NGC\,2362 was 
observed with \emph{Chandra X-Ray Observatory} on UT date 2003 December (Obs. 
ID 4469). These data have been recently reported by Damiani et al. (2005). The
dataset was filtered using the lightcurve in the 0.5 to 8 ~keV band ACIS-S CCD.
The CIAO {\sc lc\_clean} tool was used to remove flare periods. The total 
exposure time, after removing periods containing flares, is 92.7~ks. The CIAO 
{\sc celldetect} source detection routine was then used on the level 2 event 
data to produce a list of point sources. Cell sizes between 4 pixels and 8 
pixels were used. We have detected 231 X-ray point sources in the ACIS CCD in 
the range between 0.5 to 8.0 keV band.  

The ACIS field of view in these observations covers 56\% of the field 
in our UBVR$_C$I$_C$ observations. Matching with the 2MASS coordinates 
results in the identification of optical counterparts for 152 sources. Of 
these, 127 fall  in the region of the $V$, $(B-V)$ and $(V-I)$ color-magnitude 
diagrams occupied by the cluster members.

Finally, a recently published photometric and spectroscopic study of the 
cluster (D05), includes equivalent widths of H$\alpha$ in emission for 99 
stars in common with our photometric catalogue, 69 of them also with measured
equivalent width in the absorption line Li{\sc i}~6708\AA. 

In the following we consider in our analysis those stars selected as members
on the basis of our optical photometry (see below), and with values in at least
one of the data bases used. 

\section{Analysis}

\subsection{Optical Photometry}

Distance, color excess and membership to the cluster are determined with the
procedure designed by Delgado et al. (1998). Briefly, using the ZAMS line 
(Schmidt-Kaler 1982) we calculate values of color excess $E(B-V)$, visual 
absorption A$_V$=3.1$\times E(B-V)$, and absolute magnitude M$_V$ for all 
possible main sequence (MS) cluster members. A plot of $V$-A$_V$ versus M$_V$, 
allows to establish membership of evolved and unevolved MS members, and 
discard possible non members. 30 stars are selected in total as MS members. 
The mean values of color excess and distance modulus for the unevolved 
members are then used as reference values to analyse membership of the 
remaining stars. These amount to $E(B-V)=0.12\pm0.04$, $V_0-M_V=10.78\pm0.15$.
The quoted errors are the rms deviations of the mean.

In this process, we compute for every star several values of the color excess 
and distance modulus by ZAMS fitting, and also the values given by comparison 
to theoretical PMS isochrones. In absence of phenomenological reference lines 
for PMS stars, of the type of observational ZAMS, PMS isochrones are used to
assign probable membership to every star. A star is considered as member when 
color excess and distance, in respect to the isochrone, coincide inside the 
errors with the above referred reference values, in at least two CM diagrams.

In this calculation we use three sets of PMS isochrones by D'Antona \& 
Mazzitelli (1997), Palla \& Stahler (1999), and Siess, Dufour \& Forestini 
(2000) (respectively referred to in the following as D97, P99, S00)  for ages 
between 1 and 10 million years. The theoretical isochrones by D97 and P99 are
transformed to the observational CM diagrams with the calibrations by Kenyon
\& Hartmann (1995). A total of 276 stars are selected as PMS cluster members, 
in respect to at least one of the three isochrone sets used.

Because of the larger photometric errors for fainter stars, most of the PMS 
candidates are selected as members with respect to several isochrones. The 
average  value of their ages provides a formal age for every assigned PMS 
member star, and the median of all members is adopted as the age of the PMS 
cluster sequence. The resulting ages for the three sets of isochrones amount 
to 4.3$\pm$2.6, 6.0$\pm$2.4, and 5.9$\pm$2.1 Myr for D97, P99, and S00 
isochrones respectively. These values indicate coincidence of ages within the 
uncertainties and little age spread among PMS cluster members. On the other 
hand, the comparison to post-MS isochrones from the Padova group (Girardi et 
al. 2002) as plotted in Figure 1, suggests an upper limit of 4 Myr for the 
most massive MS cluster members.

The plot in Figure 1 shows post-MS isochrones for 4 and 10 Myr (Padova), and 
PMS isochrones for 1, and 10 Myr (D97, P99, S00) in the $V,(V-I)$ CM diagram. 
The stars classifed as members by any one of the fittings described above are 
marked with larger dots, while those found to be optical counterparts of 
detected X-ray sources, as well as those in common with the H$\alpha$ study 
by D05 are marked with crosses and squares, respectively. We remark that 85\% 
of the detected X-ray sources, with optical counterpart in the appropiate 
range of color and magnitudes are selected as members by our procedure, and 66
out of 80 stars with H$\alpha$ emission in the overlapping field and adequate 
magnitude range turn out to be assigned PMS members (in particular, 50 out of 
the 58 stars with both X ray detection and Ha emission). The good agreement 
between these independent criteria of both PMS nature of the stars, and 
membership to the cluster shows the reliability of the membership estimate by 
the isochrone fitting used -- by similarity to the ZAMS fitting procedure 
commonly used to estimate distances for MS stars.

The relation of ages determined from PMS and Post-MS isochrones, as well as
the possible age spread among PMS cluster members, are adressed in almost 
every investigation dealing with PMS stars in young clusters, and the 
evidences are not conclusive in either sense. In our case, as explained above,
no significant age spread is found among PMS cluster members, although 
different values are found from comparison to different PMS models. We remark 
again the comparatively younger ages determined from the D97 isochrones, and 
most interestingly,  the lower age obtained for upper MS stars than the one 
inferred for PMS cluster members. These kind of trend has been found, for 
instance, in the association Sco-Cen~OB2 by Mamajek, Meyer \& Liebert (2002), 
and would suggest later formation of most massive stars.

\subsection{Chandra-ACIS data}

The computation of X-ray luminosity from count rates for the detected sources
proceeds after the selection of the best temperature model using diagrams of 
hardness ratios. 

We compute the ratios HRA=(C$_2$-C$_1$)/(C$_2$+C$_1$) and 
HRB=(C$_3$-C$_2$)/(C$_3$+C$_2$), where C$_1$, C$_2$, and C$_3$ are the total 
counts in the bands 0.6-1.6~keV, 1.6-2.0~keV, and 2.0-8.0~keV, respectively. 
The values for our sources are compared in the HRB vs HRA diagram to a 
single temperature grid (Raymond-Smith) calculated with the {\sl Portable 
Interactive Multi-Mission Simulator} (PIMMS). This plot is shown in Figure 2.
In the range beween 0.4 and 4 
keV, the temperature model that best reproduces the distribution of our 
sources corresponds to kT=1.7\,keV, with an Hydrogen column density of 
$\rm{2.5\times 10^{21}~cm^{-2}}$, consistent with the $\rm{N_H}$ value from 
the H{\sc i} map (see 
{\tt http://heasarc.gsfc.nasa.gov/cgi-bin/Tools/w3nh/w3nh.pl}).
With these parameters, fluxes are calculated for each source 
in the ranges 0.5-2.0~keV,  2.0-8.0~keV, and the total flux in the range 
0.5-8.0~keV. To compare with the results obtained by assigning
different temperature models to PMS and MS stars, we have also calculated the 
fluxes in the same ranges for temperatures of 2.16~keV and 0.6~keV for PMS 
and MS stars, respectively (Flaccomio, Micela \& Sciortino 2003).

In Table 1 we list the data for those stars in our photometric catalogue,
which are also detected in X-rays, or included in the D05 publication. The 
table lists the identification number, and membership identification in 
columns 1 and 2: MS and PMS members are respectively denoted with 1 an 2 in 
column 2. Equatorial coordinates (Equinox 2000) in columns 3,4. Color indices
$V$, $(U-B)$, $(B-V)$, $(V-R)$, and $(V-I)$ in columns 5 to 9, $(H-K)$ 
increment with respect to the reddenning line in the $(J-H),(H-K)$ diagram in 
column 10. Equivalent widths from D05 in H$\alpha$ and Li{\sc i}~6708\AA, in
columns 11, 12. Calculated X-ray fluxes (ergs/sec.cm$^2$) in the bands 
0.6-2.0~keV, and 2.0-8.0~keV, in columns 11 and 12. And X-ray luminosity 
(calculated as the decimal logarithm of the total intrinsic flux), and 
Bolometric Luminosity, in columns 12 and 13. The color excesses and slope of 
the reddening line in the $(J-H),(H-K)$ diagram follow the reddening law from 
Cardelli, Clayton \& Mathis (1989). The intrinc total flux is calculated with 
the distance obtained from our optical photometry. The contents of Table 1 is 
accesible in electronic format.

The first evidence that stands out in the observations is the wide range of
colors and magnitudes covered by the detected X-ray sources, in particular
the considerable amount of MS stars showing this activity, as compared to
other clusters of similar characteristics, as NGC 6530 (Prisinzano et al.
2005). The deeper and more exhaustive observations of Orion, described
by Stelzer et al. (2005), also show a generalized presence of X-ray activity
among OB type stars. The authors suggest a classification in two different 
types of mechanism for the X-ray activity in their OB-type MS stars. 
Following this, the X-ray activity detected from OB stars in NGC~2362 could 
be mainly adscribed to binary companions, rather than to the presence of 
strong winds (Stelzer et al. 2005). We note however that some of these member 
stars show signs of NIR excess in the $(J-H),(H-K)$ diagram (separation from 
the reddening line larger than their error bar. See Table 1). This could be 
interpreted as a consequence of a certain amount of circumstellar material 
around these stars, which could be still in the PMS stage. 

Clear differences can be observed otherwise between the properties of X-ray 
activity in MS and PMS cluster members. Figure 3 shows a plot of logL$_X$ vs 
logL$_{Bol}$. L$_{Bol}$ is calculated with the bolometric corrections from the 
calibration by Kenyon \& Hartmann (1995). The figure shows the different 
behaviour of PMS and MS members, represented respectively as dots and crosses.
We observe that both PMS ans MS stars cover the same range in L$_X$. On the 
other hand, there is a clear correlation between both luminosities for PMS 
stars up to spectral type F (log$L_{Bol}\simeq$34.4), which vanishes for 
earlier type PMS candidate members and for MS stars. This different behaviour 
has been found as a characteristic feature in PMS versus MS stars (Preibisch 
et al. 2005). We note that most PMS members fall into the category of so 
called WTTs (weak line T-Tauri stars), as it has been shown by the H$\alpha$ 
analysis of D05. In particular, only 7 stars in our sample, have equivalent 
width of H$\alpha$ in emission larger than 10~\AA, usually adopted as the 
separating value between CTTs and WTTs (D05). 

Te spectral characteristics of both subsamples also show a different beaviour. 
In Figure 4 we plot the ratio between hard (2-8~keV) and soft (0.5-2~keV) 
fluxes, as listed in Table 1. A softening of the X-ray emission is observed 
for PMS star as L$_{Bol}$ increases. This trend again seems to dissapear for 
the MS and for the earliest type PMS candidate members. As referred above, 
the same spectral model has been used to compute fluxes for both PMS and MS
members, corresponding to a plasma at temperature of 1.6 keV. Some authors 
distinguish between both evolutive stages, assigning harder spectra to PMS 
stars than to those in the main sequence, (Flaccomio, Micela \& Sciortino 
2003). We wish to stress that the tendency shown in Figure 4 
is also apparent, and even enhanced, if we compute hard and soft fluxes with  
different models of 2.16 keV and 0.6 keV, respectively for PMS and MS stars.
The softening of X-ray activity for MS stars, even massive stars, as compared 
to PMS stars has also been established in recent works (Preibisch et al. 2005,
Stelzer et al. 2005).

As mentioned above we wish to specifically point out the behaviour of the PMS 
candidate members of the earliest spectral type in the cluster (around AF).
Although they are PMS candidates according to the optical photometry, and the
comparison to PMS isochrones, their X-ray activity show features closer to 
those from MS stars, both in the relation of L$_X$ to L$_{Bol}$, and in the 
hehaviour of their hardness ratio as well.

Finally, from the analysis of several star forming regions in a wide range of 
ages, Flaccomio et al. (2003) have concluded thax X-ray activity increases 
with age, as the envelopes and disks of stars progressively dissapear. Also 
Stassun et al. (2004) state that the emission of X-rays can be obscured or 
modulated by accretion processes, but these are not the origin of the X-ray 
activity.  

This behaviour of the X-ray activity with age is not apparent in a sample 
where all stars belong to a cluster or association, and much less in a case
like NGC2362, where the age spread is small or even absent. However, the
joint evidence from our data sources allow some check of this age effect, which
can be called more properly evolutionary effect. For 36 stars in our sample, 
there are both X ray emission, and equivalent widths of the absorption
Li{\sc i}~6708\AA ~(D05), commonly considered as a sign of PMS nature 
(Bertout 1989). The strength of this absorption (WLi) should be smaller for 
PMS stars closer to the main sequence (Mart\'\i n ~1997, Palla et al. 2005). 
When all stars can be considered to have the same age, a variation of WLi can 
still be expected simply because stars of different masses will have reached 
different PMS evolutionary stages in the same evolving time. Considering this,
we can check the dependence of L$_X$ with Wli, as an indicator of evolutionary
status in the PMS phase. We simply compute a linear fit of logL$_X$ vs $(V-I)$,
and plot the residuals versus WLi. The plot is shown in Figure 5, together 
with a median fit. The lack of a trend in this plot, and even the marginal 
indication of an increase of L$_X$ with WLi, confirms the absence of age 
spread among PMS cluster members. On the other hand, this marginal increase 
seems to indicate a tendency opposite to the general trend shown by stars in a
wide range of ages (Flaccomio et al. 2003). That is, X-ray activity is stronger
for older stars. The tendency shown in the present data allows little more
than a speculation, but if real it could be reproducing the last phase of the 
time evolution of X-ray activity from the PMS to the MS phase. In this phase,
the possible disk vanishing process and decrease of accretion activity are no 
longer present, and the evolution of X-ray emission would be entirely 
reflecting changes that take place in the stars interior.

\section{Conclusions}

The joint evidences discussed above lead to the following conclusions. The 
isochrone fitting to establish PMS membership for individual stars provides
reliable results. This follows from the agreement between membership 
asignements using isochrone fitting, and the signs of PMS nature deduced 
from observed properties of the stars, such as X-ray activity and presence of 
specific spectral features, namely H$\alpha$ emission and Li{\sc i}~6708\AA 
absorption. 

The age estimates from the three sets of PMS isochrones used provide values
wich range from 4 to 6 Myr, with the models by D'Antona \& Mazzitelli (1997), 
Siess, Dufour \& Forestini (2000), and Palla \& Stahler (1999). This 
difference can be considered as an indication of little, if any, age spread 
among the PMS cluster members. The comparison of the upper CM diagram with 
post MS isochrones from Girardi et al. (2002), indicates an upper limit for 
the MS cluster member of 4 Myr. This result seems to suggest that the most 
massive stars were formed later, when the low mass tail was already in the 
PMS evolutionary stage.

A large number of B-type MS cluster members, as compared with the findings in 
a cluster of similar properties as NGC~6530 (Prisinzano et al. 2005), are 
found to be optical counterparts of detected X-ray sources.  
A clearly distinct behaviour is observed in the properties of the X-ray
activity from PMS and MS stars. Those show a well correlated increase with of 
L$_X$ with L$_{Bol}$, and a decreasing hardness ratio, calculated as the
ratio of fluxes above and below 2 keV. The MS stars show values of L$_X$ in the
same range covered by the PMS stars, but do not show any correlation between 
both luminosities, and the suggested variation of hardness ratio with 
luminosity is also absent, whereby this parameter takes lower values than 
it does for PMS stars. These results agree with the findings and general 
properties of X ray activity in PMS stars, obtained from the analysis of 
closer star forming regions (Flaccomio et al. 2003, Preibisch et al. 2005, 
Stelzer et al. 2005).  

\acknowledgments

This work has been supported by the spanish MCYT through grants AYA2004-05395,
and AYA2003-0128. OGM acknowledges the financiation from MCYT through grant
FPI BES-2004-5044. This publication makes use of data products from the Two 
Micron All Sky Survey, which is a joint project of the University of 
Massachusetts and the Infrared Processing and Analysis Center/California 
Institute of Technology, funded by the National Aeronautics and Space 
Administration and the National Science Foundation. The work has also made use
of the BDA Open Clusters Data Base, designed and maintained by J.C. 
Mermilliod, at the Geneva Observatory, and of of the NASA ADS Abstract Service.

{}

\clearpage

\figcaption[fig1.ps] {$V,(V-I)$ Color-Magnitude diagram for NGC~2362. The 
post-MS isochrones for 4 and 10~Myr, and PMS isochrones for 1 an 10~Myr from
the three model sets used, are shown. Larger dots are photometric cluster 
members. Crosses denote X-ray detected stars, and squares, stars included in 
the H$\alpha$ catalogue by Dahm (2005) \label{fig1}}

\figcaption[fig2.ps]{Plot of ratio HRB versus HRA (see text). The grid plotted
was calculated for a Raymond-Smith model, within the ranges of temperature 
(keV) and logaritmus of column density ($cm^{-2}$) indicated by the labels. 
Triangles: PMS stars; small black stars: MS stars; big black star: mean value 
of all the data. Only stars with count number above 10 in the range 4.5-8 keV 
are used, to reduce the error of the comparison. \label{fig2}}

\figcaption[fig3.ps]{Plot of the X-ray luminosity, L$_X$ versus bolometric
luminosity, L$_{Bol}$, for the members stars 
with detected X-ray activity. Crosses denote the stars selected as MS 
members, and dots the stars selected as PMS members. \label{fig3}}

\figcaption[fig4.ps]{Hardness ratio, defined as the quotient of the hard to 
soft fluxes listed in Table 1, is plotted versus bolometric luminosity for the
member stars with detected X-ray activity. The symbols have the same meaning
as in Figure 2. \label{fig4}}

\figcaption[fig5.ps]{Plot for the member stars with both X-ray activity 
detected in the ACIS data, and equivalent width of Li6708 (Wli) from the work
of Dahm (2005). The residuals of the linear fit of L$_X$ vs $(V-I)$, 
L$_X$=31.09-0.61*$(V-I)$ (see text), are plotted versus WLi. A median fit is 
also plotted, that shows a slight positive correlation \label{fig5}}




\plotone{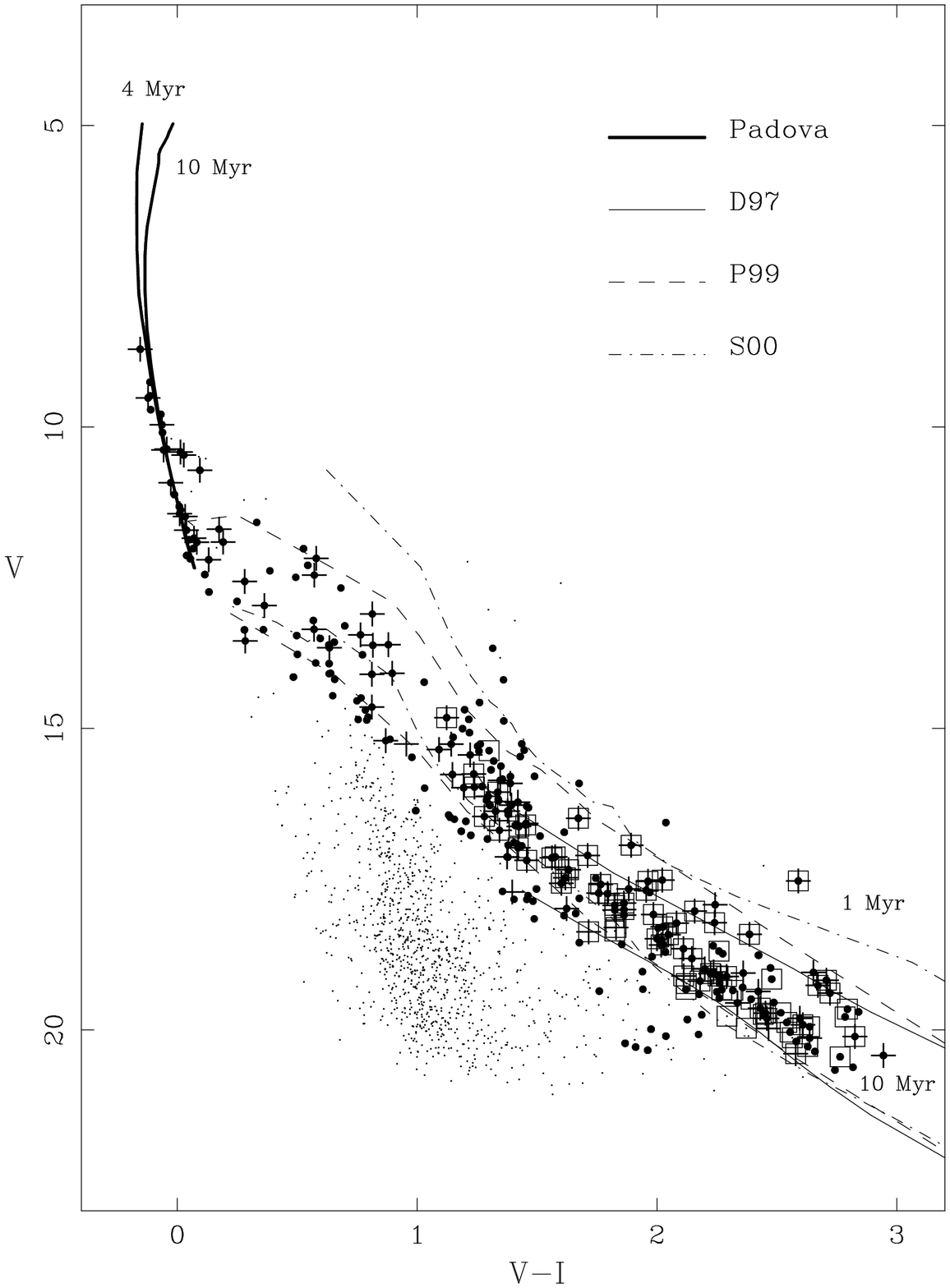}


\plotone{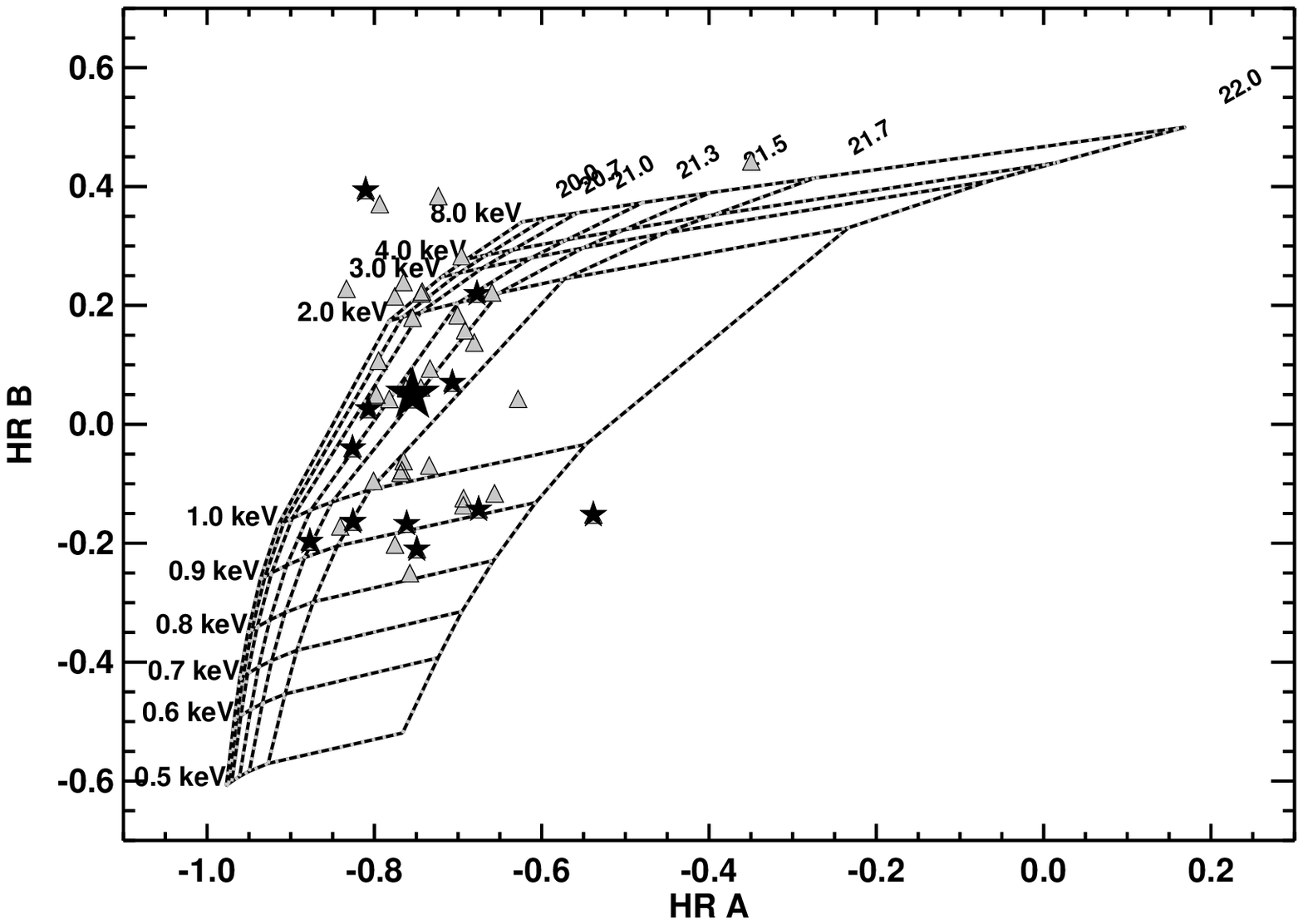}


\plotone{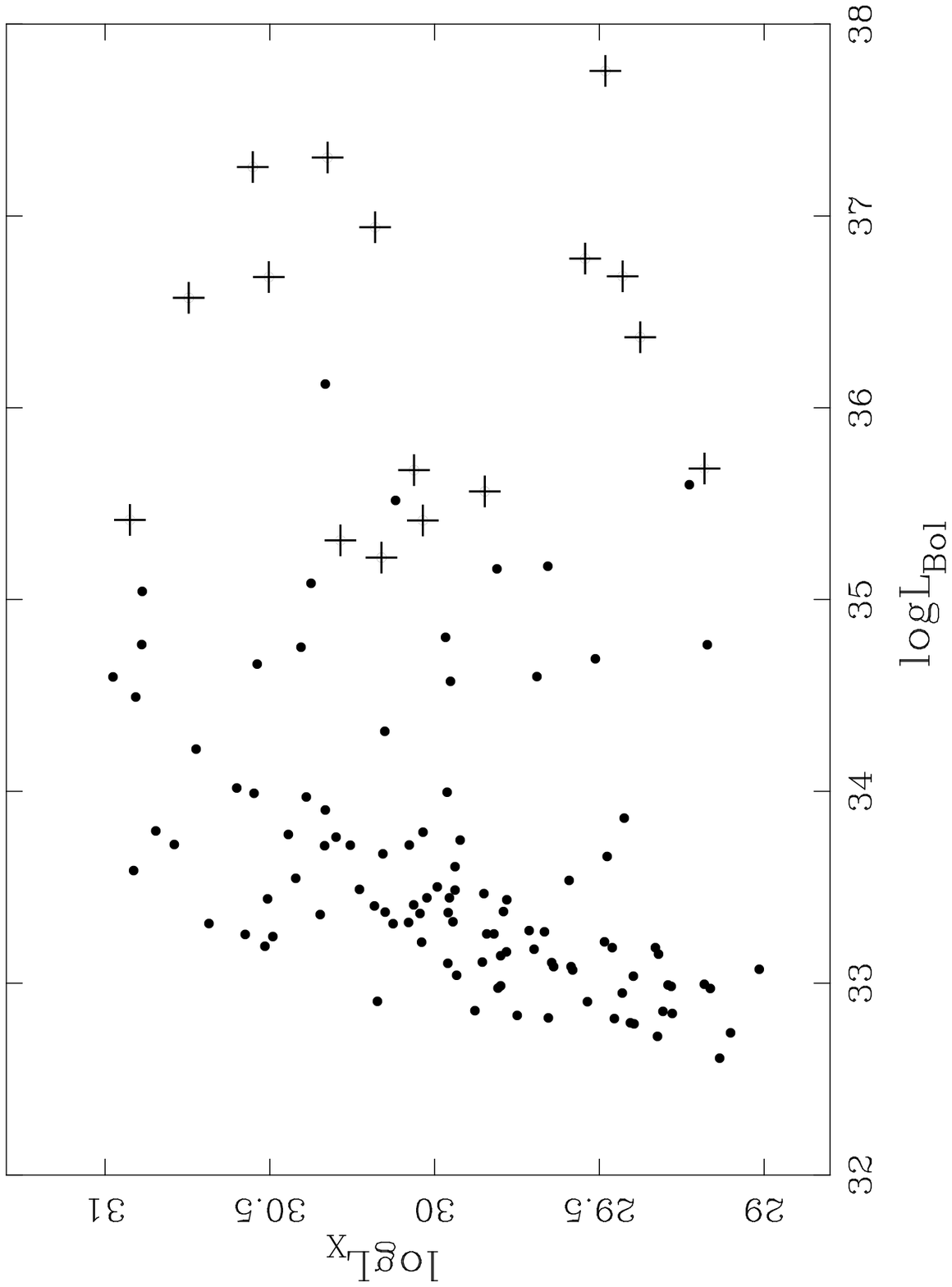}


\plotone{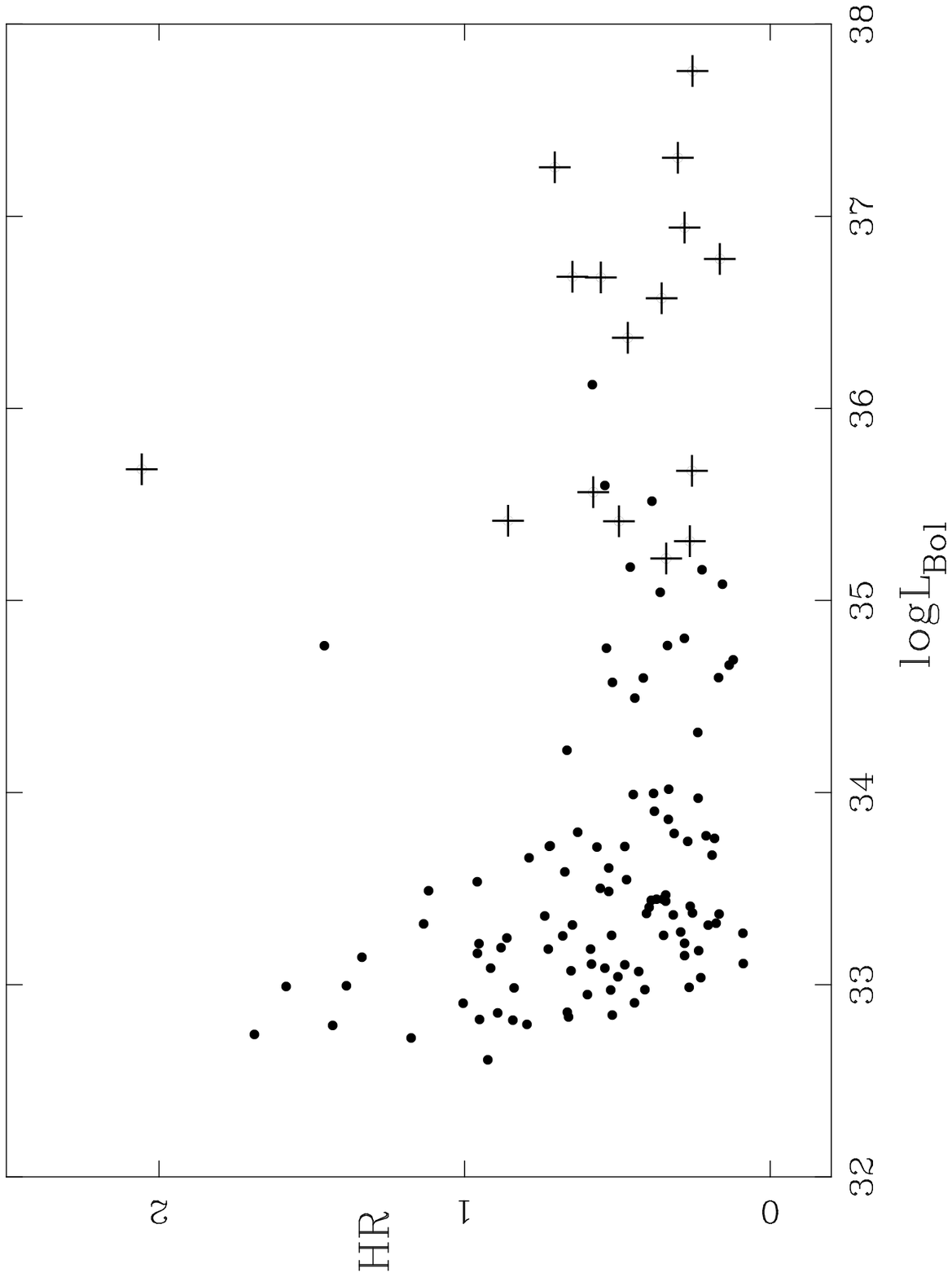}


\plotone{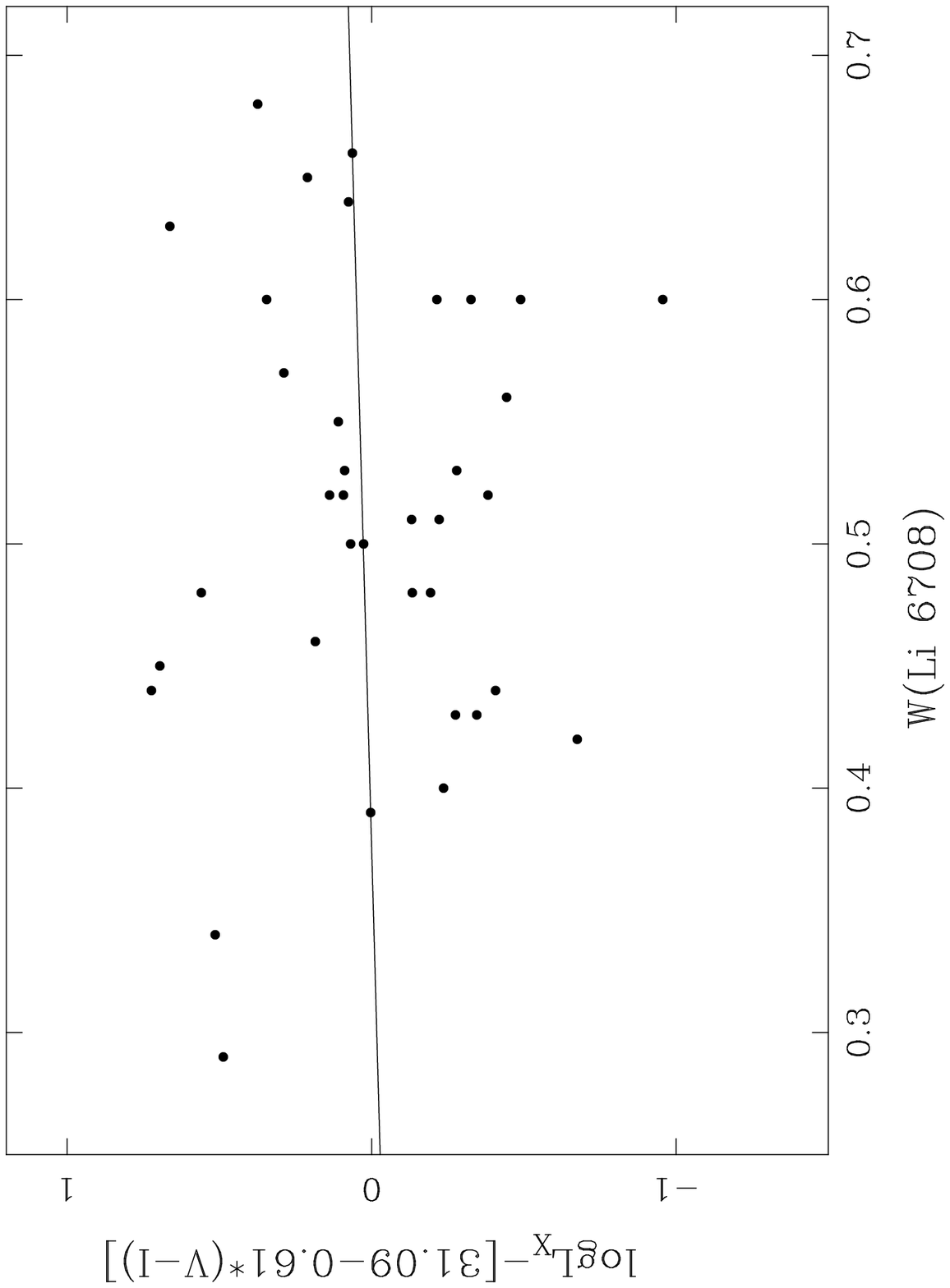}

\end{document}